\documentclass[aps,prl,reprint,amsmath,amssymb,superscriptaddress,showpacs,showkeys]{revtex4-1}
\usepackage{graphicx}
\usepackage{dcolumn}
\usepackage{bm}
\usepackage{hyperref}

\usepackage{soul} 
\usepackage{color}
\usepackage{float}

\definecolor{vs}{rgb}{0.1,0.4,0.1}                 
\newcommand{\suppress}[1]{}
\newcommand{\propose}[1]{#1}

\begin{document}


\title{Directional Fano Resonances at Light Scattering by a High Refractive Index Dielectric Sphere}


\author{Michael I. Tribelsky}
\email[E-mail: ]{tribelsky@mirea.ru}
\affiliation{Faculty of Physics, M. V. Lomonosov Moscow State University, Moscow, 119991, Russia}
\affiliation{Moscow Technological University MIREA, Moscow 119454, Russia}
\author{Jean-Michel Geffrin}
\author{Amelie Litman}
\author{Christelle Eyraud}
\affiliation{Aix-Marseille Universit\'{e}, CNRS, Centrale Marseille, Institut Fresnel UMR 7249, 13013 Marseille, France}
\author{Fernando Moreno}
\affiliation{Group of Optics, Department of Applied Physics, University of Cantabria, Spain}

\date{\today}

\begin{abstract}
In this research, we report the experimental evidence of the directional Fano resonances at the scattering of a plane, linearly polarized electromagnetic wave by a homogeneous dielectric sphere with high refractive index and low losses. We observe a typical asymmetric Fano profile for the intensity scattered in, practically, any given direction, while the overall extinction cross section remains Lorentzian. The phenomenon is originated in the interference of the selectively excited electric dipolar and quadrupolar modes. The selectivity of the excitation is achieved by the proper choice of the frequency of the incident wave.
Thanks to the scaling invariance of the Maxwell equations, in these experiments we mimic the scattering of the visible and near IR radiation by a nanoparticle made of common superconductor materials (Si, Ge, GaAs, GaP) by the equivalent scattering of a spherical particle of 18 mm in diameter in the microwave range. The theory developed to explain the experiments extends the conventional Fano approach to the case when both interfering partitions are resonant. The perfect agreement between the experiment and the theory is demonstrated.
\end{abstract}


\maketitle

\emph{Introduction.} Despite the light scattering by particles is rather an old topic, it still remains one of the most important issues of electrodynamics. Nowadays, the interest in this problem is more pronounced than ever. On one hand, it is explained by numerous applications of the phenomenon in nanotechnologies, telecommunications, medicine, biology and bioengineering, chemistry, etc.~\cite{Mishchenko}. On the other hand, it is stimulated by the discovery of very unusual properties of the light scattering. Among them, the \textit{anomalous scattering} \cite{Trib-Luk_PRL} and \textit{Fano resonances} \cite{Luk_NatMat,Mirosh_RevModPhys} should be mentioned.

The former is achieved at light scattering by particles with low dissipative losses, so that the radiative damping becomes the main effect cutting off the amplitudes of the resonant modes. For small particles, it results in narrow, well-separated resonance lines with \emph{inverted hierarchy}, so that, in contrast to the Rayleigh scattering, the partial extinction cross section \emph{increases with an increase of the order of the resonance} (diplolar, quadrupolar, etc.).
The latter occurs when the scattering wave may be presented as a sum of two partitions --- slowly depending on the incident wave frequency $\omega$ background and sharply dependent resonant. The interference between the two partitions gives rise to a typical asymmetric \mbox{$N$-shape} Fano line~\cite{Fano}.
In both cases, the scattered field amplitude exhibits a sharp dependence on the frequency of the incident wave \textit{$\omega $}. It may be effectively employed in numerous sensors, narrow-line optical filters and other devices.

However, for some applications (especially in telecommunications, information processing, optical tweezers, etc.) it is highly desirable to have the Fano profile for the intensity scattered \textit{in a given direction}. Then, a fine tuning of the frequency of the incident wave may bring the variation of the wave scattered in this direction from a very small value to its maximum, while the intensity scattered in other directions is weakly affected. As a result, one obtains \textit{tuned and controlled} redistribution of the scattered radiation in a desired direction and/or sharp spatial gradients of the scattered field.

It may be achieved by means of the \textit{directional Fano resonances}, theoretically predicted in~\cite{MIT_PRL_Fano}. These resonances are a manifestation of the conventional Fano resonances supplemented by the effects of the anomalous scattering. The role of a background partition is played by an off-resonant part of a multipole, while a sharp resonance line of another multipole represents a resonant partition. Since different multipoles have different angular dependences of the scattered radiation, at a given $\omega$ the Fano conditions may be satisfied along certain directions only. Regarding the overall scattering intensity, its profile remains symmetric Lorentzian.

Despite the apparent importance of the phenomenon, its experimental evidence has not been been obtained for quite a while. The point is that the initial theory was developed in \suppress{Ref.~}\cite{MIT_PRL_Fano} for metal particles.  Most metals are very lossy at the optical frequencies. For them, the discussed effect is suppressed by dissipation. Those which have low losses (potassium, sodium, etc.) are chemically aggressive and hardly may have any interest for practical applications.

At last, the experimental observation of the directional Fano resonances has been reported in \propose{a} recent paper~\cite{ACS_Nano_Dir_Fano} for a ``metamolecules" consisting of silicon nanospheres. {However, a number \propose{of} important questions related to intrinsic, fundamental properties of the directional Fano resonances at the wave scattering by the simplest obstacle, namely a homogeneous sphere, still remain unresolved. These questions are addressed in the present Letter.

Specifically, we inspect the scattering by a dielectric sphere with high refractive index (HRI) and low losses. The selected range of the problem parameters makes it possible to excite selectively the electric dipolar and quadrupolar modes, whose interference produces the directional Fano resonances. We measure the lineshape of the scattered intensity in a given direction and obtain \emph{pronounced typical} Fano profiles for, practically, \emph{any} scattering angle. We also measure the lineshape of the net extinction cross section (which in our case is very close to the scattering one owing to the weak dissipation) and show that in the frequency range, where the directional intensities exhibit the Fano profiles, the former is Lorentzian. Finally, analyzing the exact Mie solution we explain all the obtained experimental results analytically.} 

The scaling invariance of the Maxwell equations guarantees the identity of the scattering patterns for two geometrically similar obstacles, provided they have the same dielectric permittivity, magnetic permeability and ratio of the geometrical size to the wavelength of the incident wave. It makes possible to simulate the scattering of a visible light at the nanoscale by the corresponding scattering of microwaves. Then, the
problem becomes quite macroscopic, which allows to avoid difficulties related to precise measurements at the nanoscale. This approach has become rather a common and has been employed already in a number of works, see, e.g., \suppress{Refs.~}\cite{Vaillon2014,Moreno_NC,Belov}. In our experiments we utilize this concept of \emph{microwave analogy} to mimic the light scattering by a nanosphere made of a common semiconductor.

\begin{figure}
 \centering
   \includegraphics[width=0.5\textwidth]{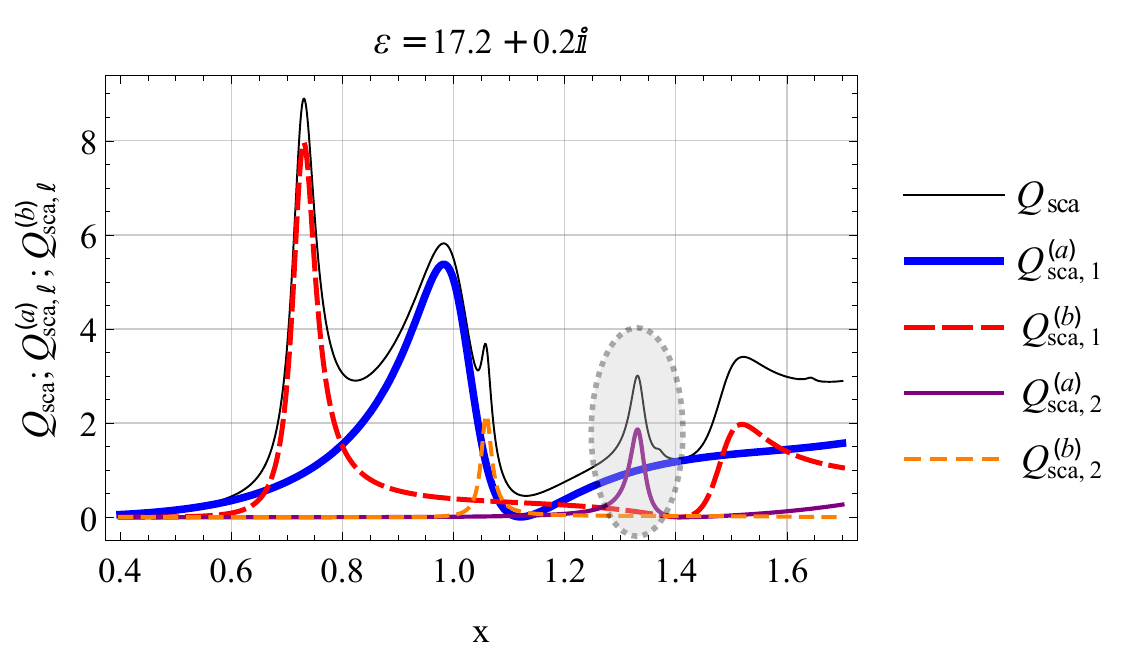}
  \caption{
  (color online) The overall scattering efficiency $Q_{\rm sca}$ and two first (dipolar and quadrupolar) partial efficiencies at \mbox{$\varepsilon = 17.2 + 0.2i$} \propose{according to the size parameter $x$}.}\label{fig:F1}
\end{figure}

\emph{The problem formulation.} The scattering of a plane, linearly polarized electromagnetic wave by a spatially uniform spherical particle is an exactly solvable problem, described by the well known Mie solution~\cite{B&W,Bohren_book}.
According to it, the scattered field is presented as an infinite series of partial waves (multipoles), such as dipole, quadrupole, etc., while each partial wave itself is a sum of the so-called electric \propose{mode} and magnetic \propose{mode}, respectively. Thus, the scattering efficiency ${Q}_{\rm sca}$ (the dimensionless ratio of the scattering cross section $C_{\rm sca}$ to the geometric one $\pi R^2$, where $R$ stands for the radius of the sphere) equals:
\begin{eqnarray*}
Q_{{\rm sca}} =\sum _{\ell =1}^{\infty }\left(Q_{{\rm sac},\ell }^{(a)} +Q_{{\rm sac},\ell }^{(b)} \right) .
\end{eqnarray*}
Here superscripts (\textit{a}) and (\textit{b}) designate the electric and magnetic contributions, respectively. Each partial efficiency, regarded as a function of $\omega$ has an infinite sequence of the Mie resonances.

The dramatic difference between these resonances for a particle with refractive index of the order of unity and those with high refractive index (HRI) is that, in the former, the resonance lines for different multipoles overlap substantially. In the latter, the overlap is much weaker. In our experiments, a homogeneous sphere with \propose{a} 18 mm diameter made from a special ceramic is employed~\cite{eccosorb}. At the working frequencies varying from 4 to 8 GHz, its complex permittivity, estimated thanks to the procedure described in \cite{Eyraud2015}, may be regarded as a constant: \mbox{$\varepsilon = \varepsilon' + i\varepsilon''=17.2 + 0.2i$}. \propose{This is} \propose{a typical value} for common semiconductors (Si, Ge, GaAs, GaP) in the visible and near IR ranges of the spectrum \cite{Forouhi1988}. The corresponding scattering efficiencies, calculated according to the exact Mie solution, are presented in Fig.~\ref{fig:F1}, where $x = kR$ is the \emph{size parameter}, $k = 2\pi/\lambda$ and $\lambda$ stands for the wavenumber of the incident wave in a vacuum.

The weak overlap of the resonance lines makes it possible to excite desired modes selectively. In our recent study~\cite{We_SciRep}, we have taken this opportunity to create tunable scattering diagrams. Here we will focus on the area, marked in Fig.~\ref{fig:F1} with an oval, where just the two modes: electric dipolar and quadrupolar make the key contribution to the overall scattering. It is important that while at the edges of the marked area, the quadrupolar efficiency is smaller than the dipolar one, in the middle of it, the case is opposite. In addition, the dipolar mode in this area is slowly varying (the background partition), while the quadrupolar mode varies sharply (the resonance partition). Such a behavior provides all the prerequisites for the manifestation of the directional Fano resonances~\cite{MIT_PRL_Fano}.

%

\emph{Experimental setup and results}. %
For the experimental study of the phenomenon, we employ the bistatic facility in the anechoic chamber of the Centre Commun de Ressources en Microondes (CCRM). Our measurement protocol is rather common in the radar cross section experiments, with the measurement of a reference target, a background subtraction and a software time gating. \propose{The measurement are performed with the emitting and receiving antennas both located in the azimuthal plane. These two horn antennas are linearly polarized and  two polarization cases are measured, with either the emitter and the receiver polarization vectors parallel (P) to the scattering (azimuthal) plane or perpendicular (S) to it}. A more detailed description of the facility and of its performances can be found in \cite{Geffrin2009, Moreno_NC, Geffrin2015, Eyraud2015}.
Such performances are rendered possible thanks to the two main items that should be pointed out in our measurement protocol.

\begin{figure}
  \centering
 \includegraphics[width=0.4\textwidth]{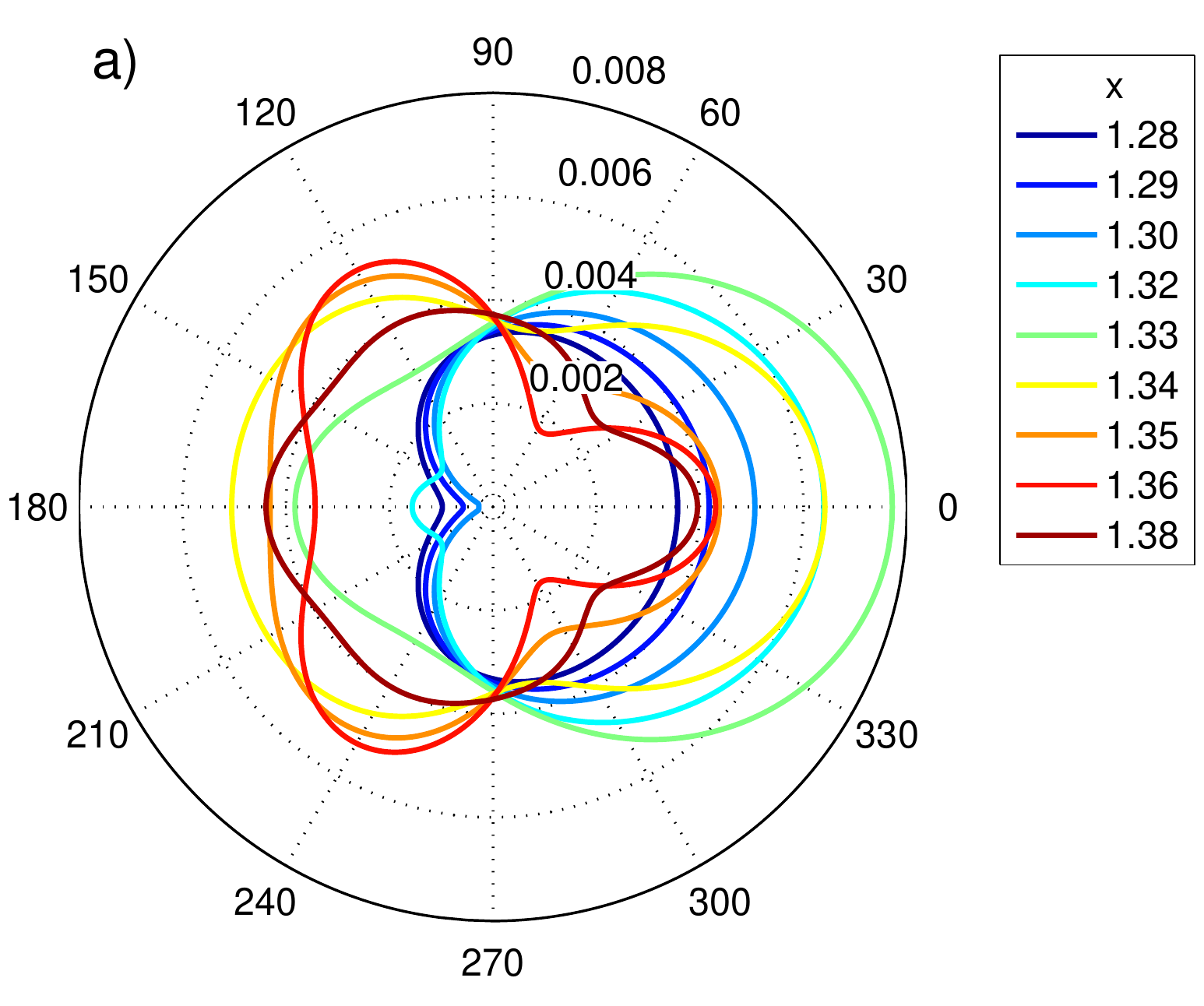}\\
 \includegraphics[width=0.4\textwidth]
  {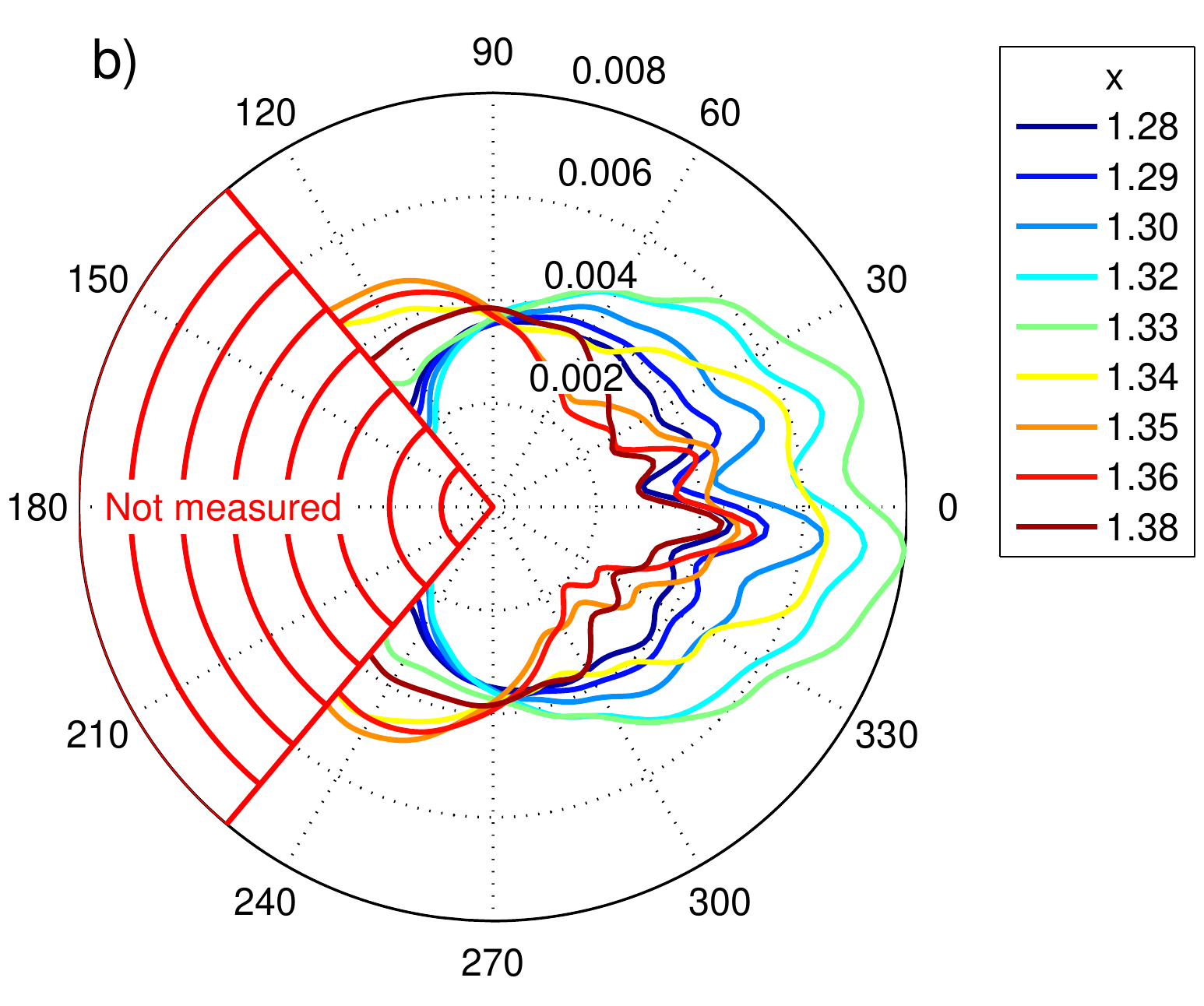}
\vspace*{-3mm}
  \caption{(color online) \propose{Angular variations of the magnitude of the scattered electric field in the S polarization case. (a) Simulation according to the exact Mie solution. (b) Experimentally measured.}
}\label{fig:diagram}
 \vspace*{-3mm}
\end{figure}

First, we employ the large angular range of our measurements (see Fig.~\ref{fig:diagram}) to enhance the post-processing data treatment. Since the scattered field is obtained from the difference between the two fields: {the net field measured by the receiver, when the scattering sphere is situated in the chamber, minus the incident field measured, when the sphere is removed,} the result is very sensitive to any kind of disturbances (e.g., the drifts, which necessarily appear during the measurements, etc.). To compensate them, we take the advantage of the angular spectral properties of the scattered field \cite{Bucci1987} in the post-processing step \cite{Eyraud2006}.

Second, as our measurements are referenced using a perfectly known target, they are calibrated so, that the presented values are all \emph{quantitative}, obtained in the \emph{dimensional units}, {in contrast to the conventional experiments in this field, whose results are presented in \emph{arbitrary units}. It makes possible to compare our measurements with the calculations based on the exact Mie solution \emph{quantitatively}, while in the conventional approach only the lineshapes may be compared.}

\begin{figure}
  \centering
  \includegraphics[width=0.45\textwidth]{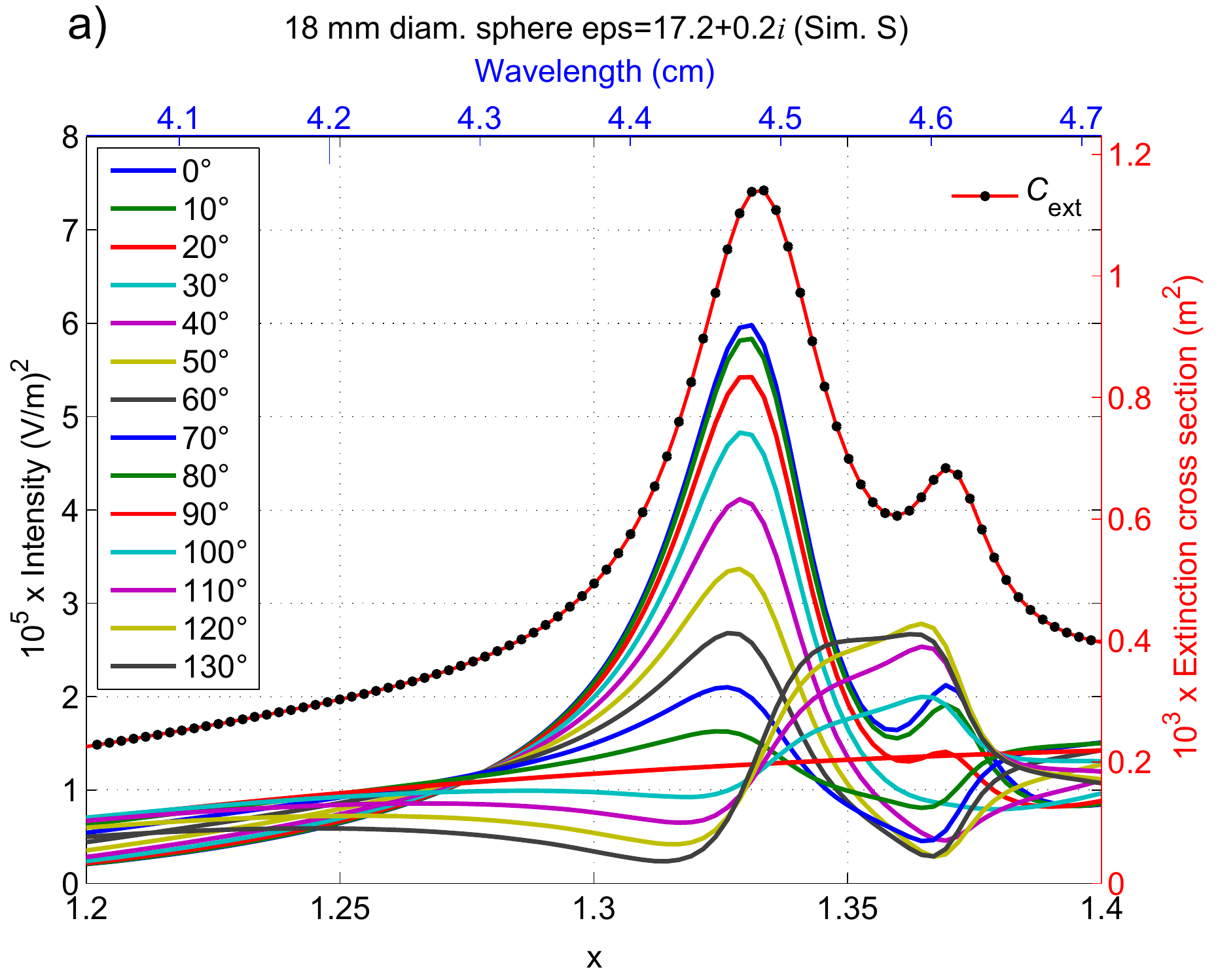}\\
  \includegraphics[width=0.45\textwidth]{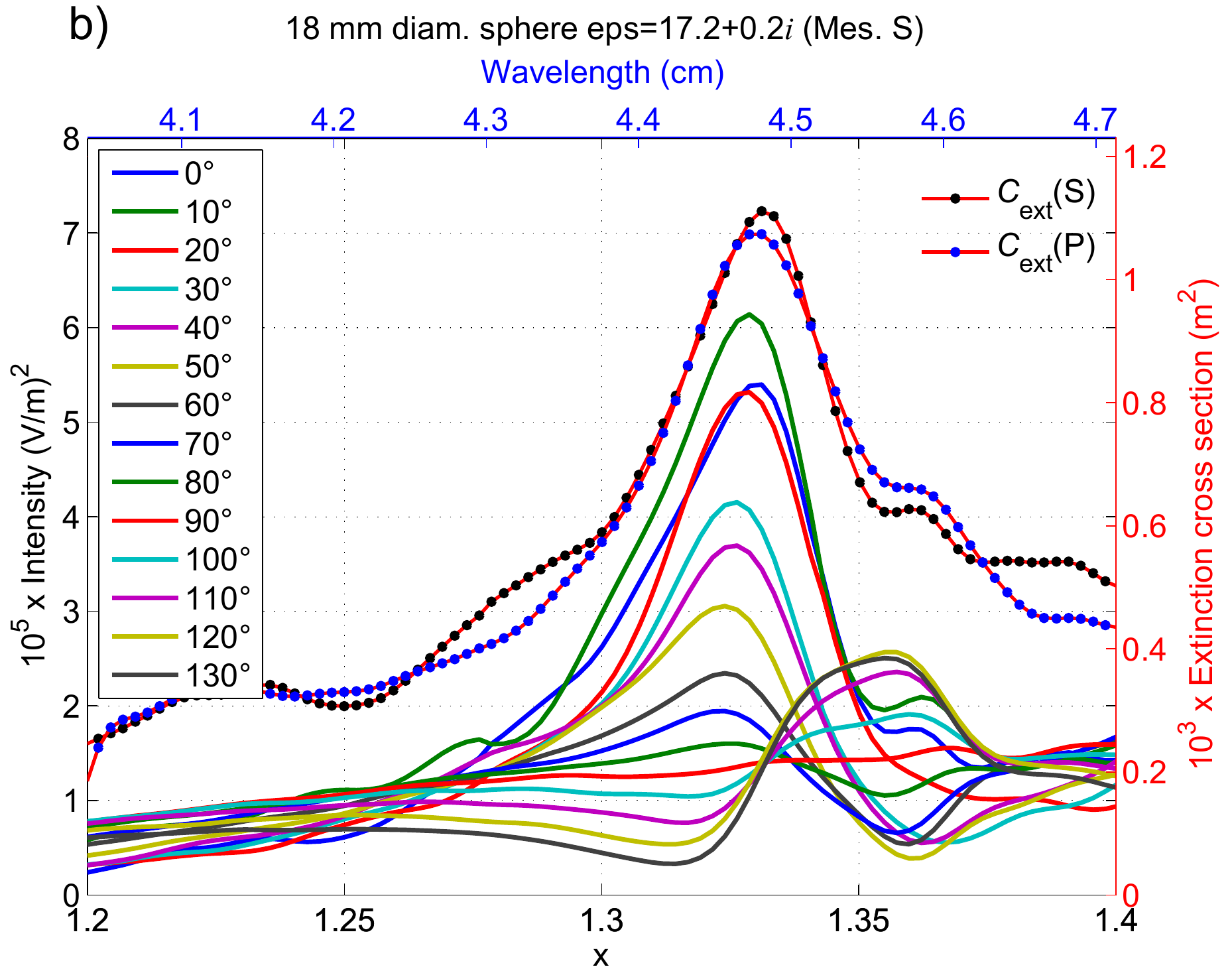}
  \caption{(color online) {The intensity of the wave in the S polarization scattered by the sphere in the azimuthal plane along the angle $\theta$, [$I_{_{\rm S }}(x,\theta)$] and the extinction cross section $C_{\rm ext}(x)$.
(a) Simulation (Sim.) according to the exact Mie solution. (b) Experimentally measured (Mes.); $C_{\rm ext}(x)$ is measured for both polarizations independently}. An increase in $\theta$ results in a monotonic decrease of the local maximum in the vicinity of $x=1.32$ until at $\theta = 90^\circ$ it vanishes. Further increase in $\theta$ transforms it into a local minimum, whose value continues to decrease with an increase in $\theta$. The case with the local minimum in the vicinity of $x=1.36$ is opposite.
\vspace*{-5mm}
}\label{fig:F2}
\end{figure}

Once \propose{the calibration and} the post-processing treatment of the complex measured scattered field have been done, the extinction cross section may be obtained by means of the optical theorem~\cite{B&W}. To this end, we \propose{extract} from the amplitude and phase of the electric field \propose{its value in the exact forward direction} \cite{Larsson2009}.  \propose{The forward zone is one of the most difficult angular range, if the measurement accuracy is a concern: In that range, the receiving antenna is blinded by the primal irradiation while the scattering field amplitude is very small. In the $\pm 20^{\circ}$ forward zone and in the selected range of size parameter, the magnitude of the scattered field is on average only $1\%$ of the actual measured field}.

The results of these measurements and their comparison with the Mie solution are shown in Fig~\ref{fig:F2}. {The extinction cross section may be independently extracted from the forward scattering with any of the two polarizations. We have over-plotted the two independently extracted values, demonstrating the very good reproducibility of our measurements.}
A pronounced set of the Fano profiles with various values of the asymmetry parameter as well as the excellent \emph{quantitative} agreement between the theory and experiment are seen straightforwardly. Note that the extinction cross section has almost perfect Lorentzian shape. A small secondary maximum \propose{noticeable} at $x \approx 1.37$ corresponds to the resonance excitation of the magnetic quadrupolar mode and has nothing to do with the discussed matter.

\emph{Discussion.}
To understand the \propose{obtained} results, we have to remember that, in the specified range of the size parameter, just the two modes: the electric dipolar and the quadrupolar one, make the key contribution to the scattered intensity. In this case, in the far zone, the intensity, scattered at a given angle \textit{$\theta $} and polarized perpendicular to the scattering plane is given by the following expression~\cite{B&W}:
\begin{eqnarray}
 I_{_{\rm S }} & \approx & 
  \left(\frac{\lambda }{2\pi r} \right)^{2}|{}^{e} B_{1}|^2 |1 + f(x)\cos\theta|^2, \label{Electric1Plus2}
\end{eqnarray}
where $r$ is the distance from the scatterer to the observation point in the spherical coordinate frame, whose origin coincides with \propose{the one of} the scattering sphere, $f(x)$ stands for $-3i(^eB_2/^eB_1)$ and \propose{the} complex quantities ${}^{e} B_{1,2} $ (the electric dipolar and quadrupolar coefficients, respectively) are expressed in terms of $x$ and \propose{the} complex refractive index $\hat{m}=n+i\kappa \equiv \sqrt{\varepsilon } $ through the Riccati-Bessel functions. The corresponding expressions are cumbersome \cite{B&W}, and will not be presented here.

\begin{figure}
  \centering
  \includegraphics[width=0.45\textwidth]{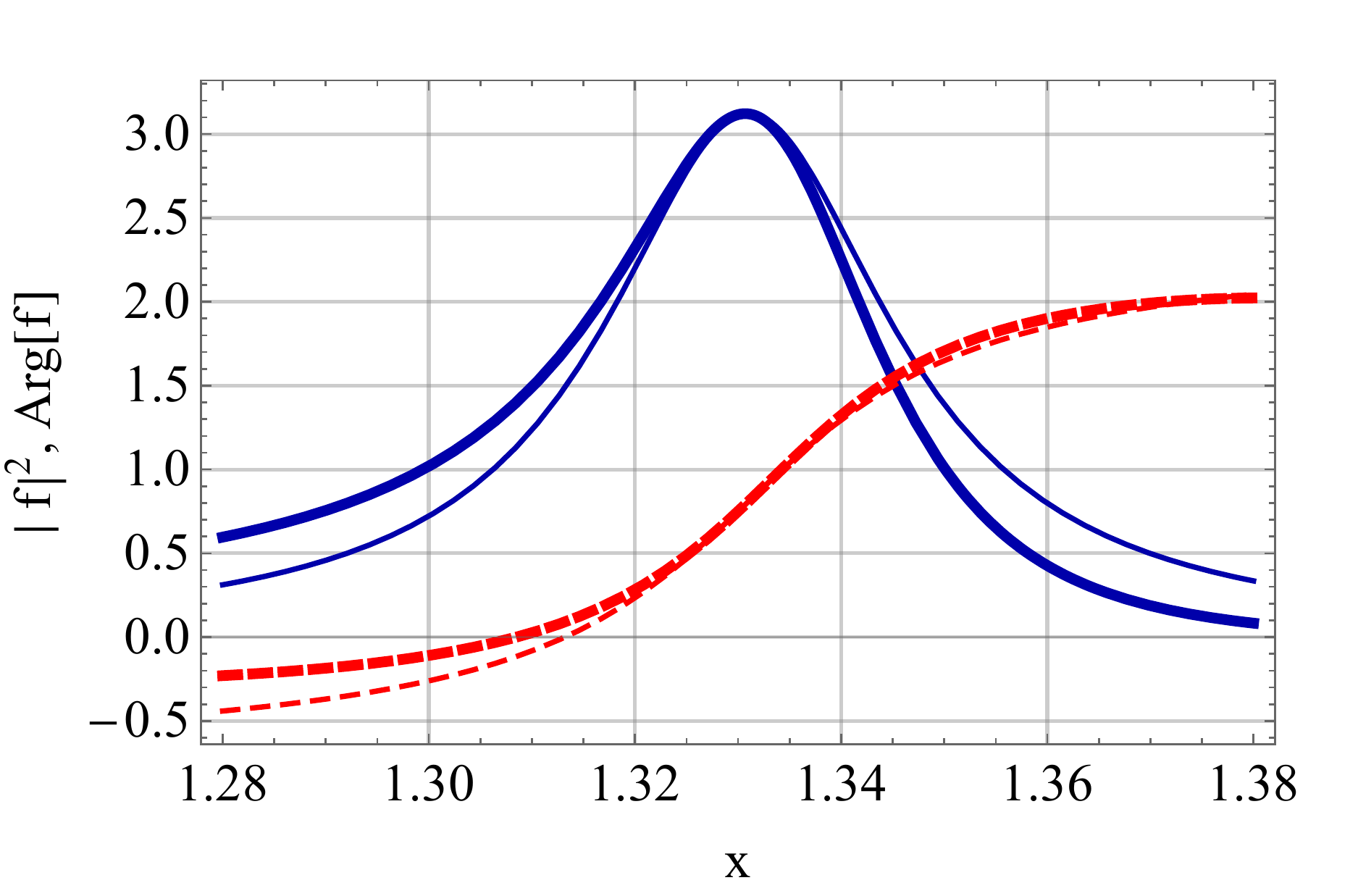}
   \vspace*{-2mm}
   \caption{(color online) Modulus (blue solid lines) and phase (red, dashed lines) of complex function $f(x) \equiv -3i(^eB_2/^eB_1)$ at $\varepsilon = 17.2 + 0.2i$ calculated according to the exact Mie solution (thick lines). Its fit by approximation \eqref{f_Lorentz} is shown with the corresponding thin lines. }\label{fig:f}
  \includegraphics[width=0.45\textwidth]{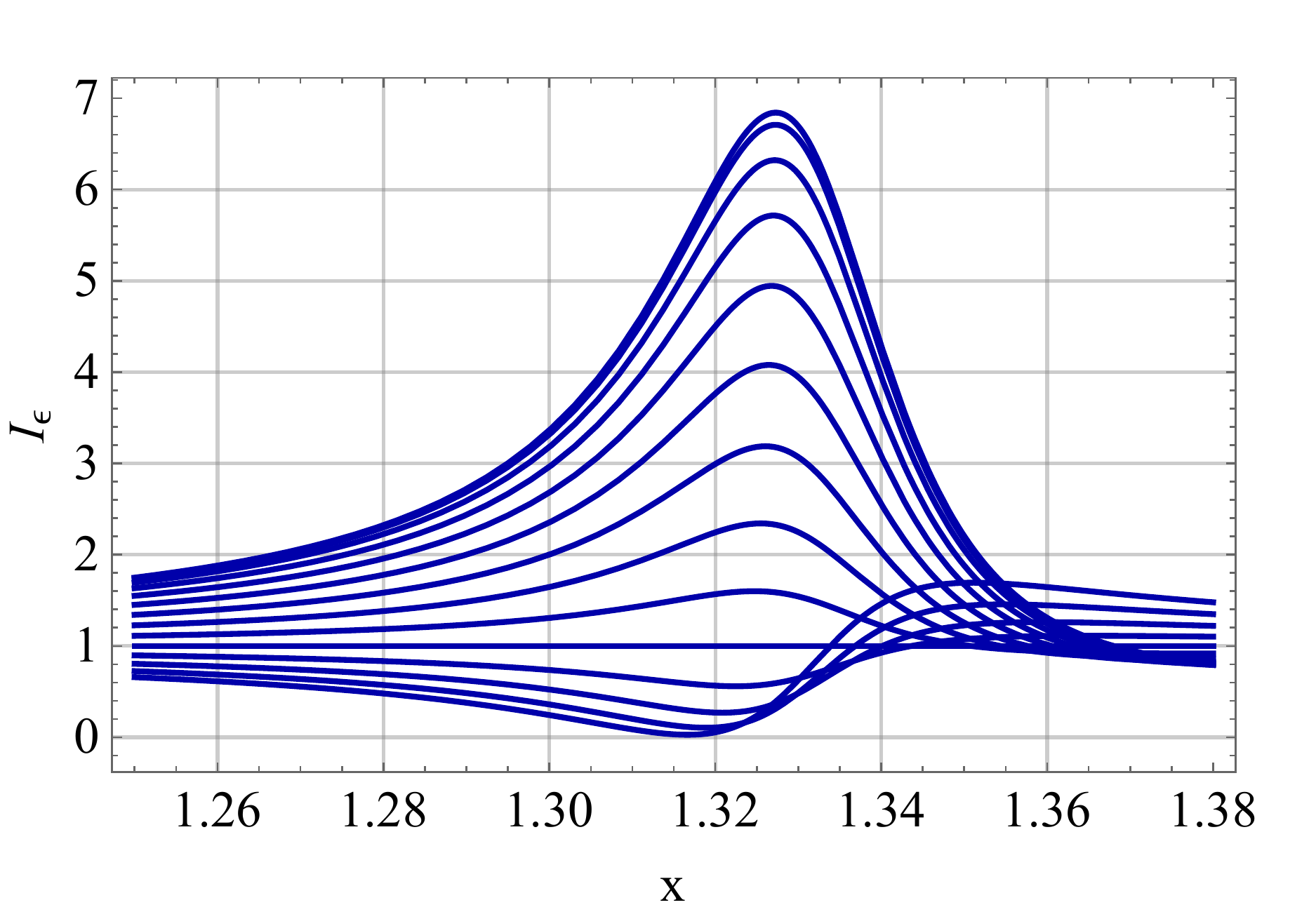}
  \caption{(color online) $I_\epsilon$ according to Eq.~\eqref{Fano} for $\theta$ varying from 0 to 130$^\circ$ with the step in 10$^\circ$ as a function of $x$, \mbox{cf. Fig.~\ref{fig:F2}.}}\label{fig:Ie}
\vspace*{-3mm}
\end{figure}

Now note, that the partial scattering efficiencies are proportional to the square of \propose{the} modula of the corresponding scattering coefficients~\cite{Bohren_book}. Then, as it follows from Fig.~\ref{fig:F1}, at $1.28 \leq x \leq 1.38$ \propose{the} quantity $|{}^{e} B_{1}|^2$ may be regarded as a constant with quite a reasonable accuracy. In this case, in the specified segment dependence, $I_{_{\rm S }}(x,\theta)$ \propose{is} practically entirely determined by \propose{the} product $f(x)\cos\theta$. On the other hand, in the discussed range of variations of $x$, \propose{the} complex quantity $f(x)$ exhibits a typical resonant behavior, see Fig.~\ref{fig:f}, and may be approximated by the expression
\begin{equation}\label{f_Lorentz}
  f(x) \approx A\frac{(\Gamma/2)e^{i\phi_0}}{(x-x_{\rm res}) + i(\Gamma/2)}.
\end{equation}

The values of the fitting parameters employed to obtain Fig.~\ref{fig:f} are as follows: $A=1.7669,\; \Gamma = 0.034$, \mbox{$\phi_0 = 2.379,\; x_{\rm res} = 1.331$.}

\propose{The s}ubstitution of Eq.~\eqref{f_Lorentz} into Eq.~\eqref{Electric1Plus2} yields the following profile for the scattered intensity:
\begin{equation}\label{Fano}
  \left(\frac{2\pi r}{\lambda |{}^{e} B_{1}(x_{\rm res})|} \right)^{2} I_{_{\rm S }} \approx I_\epsilon \equiv \frac{(\epsilon + q)^2}{\epsilon^2 + 1} + \frac{(1 + q\tan\phi_0)^2}{\epsilon^2 +1},
\end{equation}
where $\epsilon \equiv 2(x-x_{\rm res})/\Gamma,\; q \equiv A\cos\phi_0\cos\theta$.

Thus, the obtained profile is a superposition of the conventional Fano line~\cite{Mirosh_RevModPhys} (the first term in the right-hand-side of Eq.~\eqref{Fano} and the Lorentzian profile (the second term). It is important \propose{to note} that, while the positions of the extrema of the Fano profile (the minimum at $\epsilon = -q$ and maximum at $\epsilon = 1/q$) depend on the scattering angle, the position of the maximum of the Lorentzian line is fixed at $\epsilon = 0$. It means that, \propose{by} varying $\theta$, we may vary the distance between the minimum of the Fano profile and the maximum of the superimposed Lorentzian one, controlling the sharpness of the corresponding part of $I_\epsilon$. This additional way to control the lineshape may be extremely important for applications of the discussed effects in sensors and similar devices.

To illustrate the description of the actual lineshapes by \propose{the} approximation \eqref{Fano}, the corresponding profiles for $\theta$ varying from 0 to 130$^\circ$ with \propose{a} 10$^\circ$-step (the same range and sampling as those in Fig.~\ref{fig:F2}) are presented in Fig.~\ref{fig:Ie}. The values of the fitting parameters are the same as those employed to plot Fig.~\ref{fig:f}. For easier comparison of Figs.~\ref{fig:F2} and \ref{fig:Ie}, $I_\epsilon$ in the latter is plotted as a function of $x$. {The slower decay of the right wins of the lines in Fig.~\ref{fig:Ie} relative to that in Fig.~\ref{fig:F2} is explained by the departure in this range of approximation \eqref{f_Lorentz} for $f(x)$ from the actual profile, obtained from the Mie solution. 
Apart \propose{from} this point, the agreement between the two sets of figures is quite impressive.

As for the shape of the net scattering cross section, the latter is a sum of positive contributions of every partial ones. In t\propose{u}rn, the partial cross sections are proportional to the squares of the modula of the corresponding scattering coefficients~\cite{Bohren_book}. In the discussed range of variations of the size parameter, just two coefficients: $^eB_1$ and $^eB_2$ make the overwhelming contribution to the net cross section. Moreover, we have shown that, in this range, $|^eB_1|^2$ may be regarded as a constant, while $|^eB_1|^2$ has a Lorentzian shape, see Figs \ref{fig:F1}, \ref{fig:f}. These arguments guarantee the Lorentzian shape of the net scattering cross section in the discussed domain of variations of $x$. The same is true for the extinction cross section since the difference between the two is negligibly small owing to the weak dissipation.

\emph{Conclusions}. In this Letter, (i) thanks to the microwave analogy principle, we have mimicked the scattering of a plane linearly polarized light in the visible and near IR range by a homogeneous nanosphere made of a common semiconductor with the scattering of a macroscopic sphere made of a special ceramic. 
(ii). This approach makes it possible to obtain detailed experimental results on the dependences of the scattering intensity on the scattering angle and the size parameter as well as the lineshape for the net extinction cross section. (iii) We have found such a range of the size parameter where just the two modes: electric dipolar and quadrupolar make the overwhelming contribution to the net scattering cross section and shown that the interference of these two partial waves gives rise to the \emph{directional Fano resonances}. At the same time the lineshapes of the net extinction and scattering cross sections in this range remain Lorentzian. (iv) We have obtained a simple analytical expression for the lineshape of the intensity scattering along any given direction and shown that it consists of two superimposed profiles. One of them occurs the conventional Fano profile, while the other is Lorentzian. (v) For the Fano profile, the values of the extrema and their positions both depend on the scattering angle, while for the Lorentzian profile only the value of its maximum depends on this quantity, while its position is fixed. It provides additional opportunities to tailor and engineer the lineshape by varying the scattering angle. The latter may be important for practical applications of the discussed effects in sensors and analogous devices.

\emph{Acknowledgements}
This research has been supported by MICINN (Spanish Ministry of Science and Innovation, project FIS2013-45854-P). We also acknowledge the opportunity provided by the Centre Commun de
Ressources en Microonde to use its fully equipped anechoic chamber.

\end{document}